\documentstyle[12pt]{article}
\begin{document}
\thispagestyle{empty}
\begin{flushright}
CERN-TH/96-72\\
BI-TP/96-43
\end{flushright}
\vskip3cm
\centerline{\bf ON THE SUM RULE APPROACH}
\vskip0.5cm
\centerline{\bf TO QUARKONIUM-HADRON INTERACTIONS}
\vskip2cm
\centerline{
D. Kharzeev$^1$, H. Satz$^{1,2}$,
A. Syamtomov$^3$ and G. Zinovjev$^3$}
\vskip1cm
\centerline{1 Fakult\"at f\"ur Physik, Universit\"at Bielefeld, D-33501 Bielefeld,
Germany}\par
\centerline{2 Theory Division, CERN, CH-1211 Geneva, Switzerland}\par
\centerline{3 Bogolyubov Institute for Theoretical Physics, 252143
Kiev-143, Ukraine}\par

\vskip3cm
\begin{abstract}
We extend a recent sum rule calculation for inelastic quarkonium-hadron
interactions to realistic parton distribution functions; we also include
finite target-mass corrections. Both modifications are shown to have no
significant effect on the resulting cross section behaviour but the
performed analysis gives useful insights on the sum-rule
approach in general.
\end{abstract}
\vskip 4cm
\begin{flushleft}
CERN-TH/96-72\\
BI-TP/96-43\\
May 1996
\end{flushleft}
\pagebreak
In a recent study \cite{KS3}, the cross section $\sigma_{N\Phi}(s)$ for
inelastic quarkoni\-um-nucleon collisions was calculated from sum rules
established on the basis of the operator-product expansion
\cite{Peskin}--\cite{Kaidalov}; here $\Phi$ denotes a quarkonium ground state
(J/$\psi$ or $\Upsilon$) and $s$ the squared center of mass collision energy.
In order to simplify the calculations a number of approximations
was made in this investigation. The aim of the present work is to obtain
$\sigma_{h\Phi}(s)$ avoiding such approximations and to clarify some
aspects of the sum rule approach in general.
\par
In the formalism developed to describe the
interaction of hadrons with heavy $q\bar{q}$ mesons
\cite{Peskin}, \cite{Gottfried}, \cite{Voloshin},
the large quark mass $m_Q$ allows one to write down the corresponding
Born amplitude in the form of operator product expansion (OPE). The effects
governed by short-distance physics, including the structure of heavy
quarkonium state, factorize into coefficient functions, $C_n$, while
the thorough dependence on soft structure of hadron is determined by
low-energy matrix elements of gauge-field local operators, ${\cal O}_n$,
renorma\-li\-zed at quarkonium mass scale $M_\Phi$,
$$
{\cal M}_{Born}=\sum_{n=2,4,...}^\infty C_n\,
\langle{\cal O}_n\rangle.
\eqno(1)
$$
The operators entering (1) are ordered by their dimensions.
If the characteristic distances of internal $q\bar{q}$ dynamics are
small enough as compared to scale of QCD-vacuum fluctuations inside the
hadron, only the terms of lowest dimension contribute.
In the framework of the multipole expansion formalism \cite{Gottfried},
\cite{Voloshin}, the lowest dimension terms correspond to the dipole
field contribution.
For the sake of simplicity, we have dropped in (1) spin indices which
otherwise mark the amplitudes describing the forward scattering
of heavy quarkonia with different polarizations. If we restrict
ourselves to the case of spin-averaged interactions, then Lorentz
indices may appear only in the following combination:
$$
{\cal O}_n={1\over M_\Phi^n}K_{\mu_{1}}\ldots K_{\mu_{n}}\theta^G_{\mu_1
\ldots \mu_n},
\eqno(2)
$$
where
$$
\theta^G_{\mu_1\ldots \mu_n}=i^{n-2}\left[G_{\mu_1\nu}{\cal D}_{\mu_2}
\ldots{\cal D}_{\mu_{n-1}}G_{\mu_n\nu}\right](symmetrized)-traces. \label{3}
\eqno(3)
$$
Here $K_\mu$ is the quarkonium momentum, $G$ is the gluon field operator
and ${\cal D}$ means the covariant derivative.

The target matrix elements of these operators are completely specified
by pheno\-me\-nological parameters $A_n$ of dimension defined as
twist and the spin-averaged tensor structure,
$$
\langle p|\theta^G_{\mu_1\ldots \mu_n}|p\rangle=A_n
\left(p^{\mu_1}\ldots p^{\mu_n}-traces\right),
\eqno(4)
$$
thus forming into traceless, symmetrical rank-$n$ tensors.
Substituting (4) and (2) into (1) we come to schematic expression
$$
{\cal M}_{\phi p}=\sum_{n=2,4,...}^\infty C_n\,
A_n{{K_{\mu_1}\ldots K_{\mu_n}}\over{M_\Phi^n}}\Pi^{\mu_1\ldots \mu_n},
\eqno(5)
$$
where for convenience we used the notation $\Pi^{\mu_1\ldots \mu_n}=
\left(p^{\mu_1}\ldots p^{\mu_n}-traces\right)$.

At this point, we comment briefly on the applicability of the analysis
to actual physical states. For sufficiently heavy constituent quarks,
the state $\Phi$ is localized at characteristic distances
$r$ which are sufficiently small compared to non-perturbative hadronic
scale $\Lambda_{QCD}^{-1}$ to treat the system perturbatively.
On the other hand, it satisfies $r\gg m_{Q}^{-1}$, so that one can describe
the internal dynamics nonrelativistically. Although the estimates
\cite{Peskin} show that these requirements are
reliably satisfied only for quark masses exceeding $25\ GeV$, one finds
that a Coulomb-like potential approach provides a satisfactory
description of quarkonium spectroscopy (see, e.g. \cite{Spectroscopy}).
Hence we may also expect a one-gluon exchange picture to be relevant
for J/$\psi$ and $\Upsilon$ internal dynamics.
We thus consider $\Phi$ as SU(3)-Coulombic bound state
characterized by a ``Bohr" radius $a_0$
and the corresponding ``Rydberg" energy $\epsilon_0$:
$$
a_0={4\over {3m_Q\alpha_s}}, \ \ \
\epsilon_{0}=\left({{3\alpha_s}\over{4}}\right)^2{m_Q},
\eqno(6)
$$
where $\alpha_s$ is the gauge coupling constant, evaluated at a scale
$\epsilon_0$. Then the direct calculations of \cite{Peskin} in $SU(N)$
gauge theory give
$$
C_n^{(1S)}=a_0^3\epsilon_0^{2-n}d_n^{(1S)}, \ \
d_n^{(1S)}=\left({32\over N}\right)^2\sqrt{\pi}
{{\Gamma\left(n+{5\over2}\right)}\over{\Gamma(n+5)}}.
\eqno(7)
$$
for the Wilson coefficients $C_n$ of $1S$-quarkonium state in the leading
order in $1/N^2$ (analogous expressions for the $P$-states can be found
in Ref.\  \cite{DK}).

We now insert the coefficients (7) into Eq. (5). If we neglect the
trace terms in Eq.\ (4), or, equivalently, the corrections of order
of $m_N^2/\epsilon_0^2$, we recover the sum rules used in
\cite{KS3} as the basis for calculating $\sigma_{N\Phi}(s)$,
$$
\int_0^1 dy\ y^{n-2}\sqrt{1-y^2} \sigma_{N\Phi}(m_N / y) = I(n)\
\int_0^1 dx\ x^{n-2} g(x, Q^2 = \epsilon_0^2), \eqno(8)
$$
with $I(n)$ given by
$$
I(n) =  2 \pi^{3/2}\ \left(16 \over 3 \right)^2\
{{\Gamma \left(n + {5\over 2} \right)} \over {\Gamma (n + 5)}}\
\left( {4 \over {3\alpha_s}}\right)\ {1 \over m_Q^2}. \eqno(9)
$$
One obvious problem in applying these sum rules to
J/$\psi$ interactions is
that ratio $m_N^2/\epsilon_0^2\simeq 2.1$ is not actually
small and hence cannot be ignored. In other words,
the trace terms entering
the definition of twist-two operators (4) in the Wilson's ordering scheme
must be included explicitly producing the corresponding changes in
dispersion sum rules (8).

The tracelessness of the tensor $\Pi^{\mu_1\ldots \mu_n}$ (\ref{3})
means that
$$
g_{\mu_i\mu_j}\Pi^{\mu_1\ldots\mu_i\ldots\mu_j\ldots\mu_n}=0.
\eqno(10)
$$
The most general structure of $\Pi^{\mu_1\ldots \mu_n}$ is
well known \cite{GeorgiPolitzer}
$$
\Pi^{\mu_1\ldots \mu_n}=\sum_{j=0}^{n/2}(-1)^j{{(n-j)!}\over{2^jn!}}m_N^{2j}
\sum_{all\ permut.} \overbrace{g\ldots g}^{j}
\overbrace{p\ldots p}^{n-2j},
\eqno(11)
$$
where the second sum runs over the ${n!} / (n-2j)!(2j)!$ terms
of all possible permutations.
Introducing $\lambda=(pK/ M_\Phi)$ we can rewrite (5) as
$$
{\cal M}_{\phi p}=\sum_{n=2,4,\ldots}^\infty d_na_0^3\epsilon_0^2A_n
\sum_{j=0}^{n/2}(-1)^j{{(n-j)!}\over{4^j(n-2j)!j!}}\left(
{{m_N^2}\over{\epsilon_{0}^{2}}}\right)^{j}\left(
{{\lambda}\over{\epsilon_{0}}}\right)^{n-2j}.
\eqno(12)
$$
Changing the summation index $n\rightarrow n-2j$ and making
simple rearrangements of the coefficients, one obtains
$$
{\cal M}_{\phi p}=a_0^3\epsilon_0^2\sum_{n=0,2,\ldots}^\infty
\left({{\lambda}\over{\epsilon_{0}}}\right)^n\sum_{j=0}^{\infty}
(-1)^jd_{n+2j}A_{n+2j}{{(n+j)!}\over{4^jn!j!}}\left(
{{m_N^2}\over{\epsilon_{0}^{2}}}\right)^{j}.
\eqno(13)
$$
We now take the explicit expressions (7) for $d_n$ (the term
proportional to $d_0$ corresponds to disconnected diagrams and does
not enter the amplitude of quarkonium interactions in the OPE formulation)
and use the definition of $A_n$ as Mellin transforms \cite{Parisi}
of the gluon distribution in a proton, evaluated at the scale
$Q^2=\epsilon_0^2$,
$$
A_n=\int_0^1\, dx\, x^{n-2}g(x,Q^2=\epsilon_0^2).
\eqno(14)
$$
This leads to the final expression

$$
{\cal M}_{\phi p} = 4\sqrt{\pi}a_0^3\epsilon_0^2
\left({16\over3}\right)^2\left[\int_0^1\, dx
\sum_{n=2,4,\ldots}^\infty x^{n-2}
\left({{\lambda}\over{\epsilon_{0}}}\right)^ng(x,\epsilon_{0}^{2})\times\right.
$$
$$
\times {{\Gamma(n+{5\over2})}\over{\Gamma(n+5)}}
\,{}_3F_2\left({5\over4}+{n\over2},{7\over4}+{n\over2},1+n;
{{(5+n)}\over{2}},3+{n\over2};
-{{m_{N}^{2}}\over{4\epsilon_{0}^{2}}}x^{2}\right)
$$
$$
-\left.\frac{m_N^2}{4\epsilon_0^2}\frac{\Gamma({9\over2})}{\Gamma(7)}
\int_0^1\,dx\,g(x,\epsilon_0^2)\,{}_3F_2\left(1,{9\over4},{11\over4};
{7\over2},4;-{{m_{N}^{2}}\over{4\epsilon_{0}^{2}}}x^{2}\right)\right].
\eqno(15)
$$
Introducing the variable $y=m_N/\lambda$, we get
$$
\int_0^1\,dy\,y^{n-2}(1-y^2)^{1/2}
\sigma_{N\Phi}(m_N/y) =
$$
$$
I(n)\int_0^1\,dx\,x^{n-2}
g(x,\epsilon_{0}^{2})\,{}_3F_2\left({5\over4}+{n\over2},
{7\over4}+{n\over2},1+n;{{(5+n)}\over{2}},3+{n\over2};
-{{m_{N}^{2}}\over{4\epsilon_{0}^{2}}}x^{2}\right).
\eqno(16)
$$
Comparing Eqs. (8) and (16), we see that the inclusion of finite mass
corrections effectively modifies the gluon distribution function -- in the
r.h.s. one should integrate including confluent hypergeometric function
${}_3F_2(\ldots)$ as a weight.
In Eq. (16), $n=2, 3, ...$, so that in order to determine the cross section,
it appears that we have to solve an infinite set of equations. However,
usually the application of the sum rules to deep-inelastic lepton-hadron
scattering \cite{Rujula} and heavy meson photoproduction \cite{NSVZ}
had been limited to the values of $n\leq 5-6$ [1-5]: it was
 expected that perturbative calculation of the coefficient functions
breaks down for large $n$.
The use of Eq.(16) can effectively extend the range of applicability
of the sum rule approach, and the problem of relevant
$n$-interval should be re-examined.
Indeed, the additional weight function
${}_3F_2$ in the r.h.s. of (16) decreases for $x\rightarrow 1$
faster for larger $n$, making the sum rules more sensitive to the
behaviour of parton distribution function (PDF) at lower values of $x$.
This makes the reliability
 region for gluonic distribution function
moments effectively wider.
In what follows we study the sum rules for values $n\leq 10$.

It is evident from Eq.\ (16) that the behaviour of the cross section at high
energy is particularly sensitive to the small $x$ region, while the
threshold behaviour probes the large $x$ region.
New data from deep inelastic scattering have led to different PDF
parametrizations; hence one has to check how well the sum rules,
including target mass corrections as well, are satisfied for different
PDF choices. In \cite{KS3}, the simple schematic form
$$
g(x)={\rm const.}(k+1)(1-x)^4
\eqno(17)
$$
was used. Here, we have considered in addition the MRS $D\_'$ and the new
MRS $H$ \cite{Martin} parametrizations; the latter takes into account the
small-$x$ behavior observed at HERA \cite{HERA}. In all cases, the left hand
and right hand sides of Eq. (16) agree for $2\leq n\leq 10$ within better
than $1 \%$. The sum rules are thus well satisfied for all three PDF forms
used.

Next we want to check what effect  the finite mass corrections have.
In Fig.\ 1, we therefore show the ratio of the cross section including the
target mass corrections (i.e. the solution of Eq.(16)) to the same
calculation without them (from Eq.(8)).
It is seen that the inclusion of target mass corrections changes the
resulting cross section above the threshold ($\sqrt{s}>5\ GeV$) by
less than a factor three.
\par
Finally we want to see if and how the cross sections obtained as solution of
Eq. (16) vary for different PDF forms; we have therefore calculated
$\sigma_{N\Phi}(s)$ also for MRS $D_{-}^{\prime}$ and MRS $H$. The resulting
cross sections, divided by the corresponding form obtained using PDF (17),
are shown in Fig. 2; here all cross sections include target mass corrections.
The increase of the two MRS forms at high energy,
relative to that using Eq. (17), is due to the small $x$ increase in the more
realistic PDF's.

In summary, we would like to stress that including finite target mass
corrections justifies the use of OPE close to the
threshold and meanwhile does not change significantly the behaviour of
quarkonium-hadron cross section at high energies. Besides, target mass
corrections are important in providing the basis for future investigations
of the role of higher terms of the multipole expansion and for a
description of the entire $x$ region by the sum-rule method.
\vskip0.5cm
\centerline{\bf Acknowledgements:}
\vskip0.3cm
This work was supported by the GSI under grant BISAT (D. Kh.) and by INTAS
under grant 3941 (A. S. and G. Z.).

\vskip2cm
\bigskip
\centerline{\bf Figure Captions:}

\bigskip
\noindent
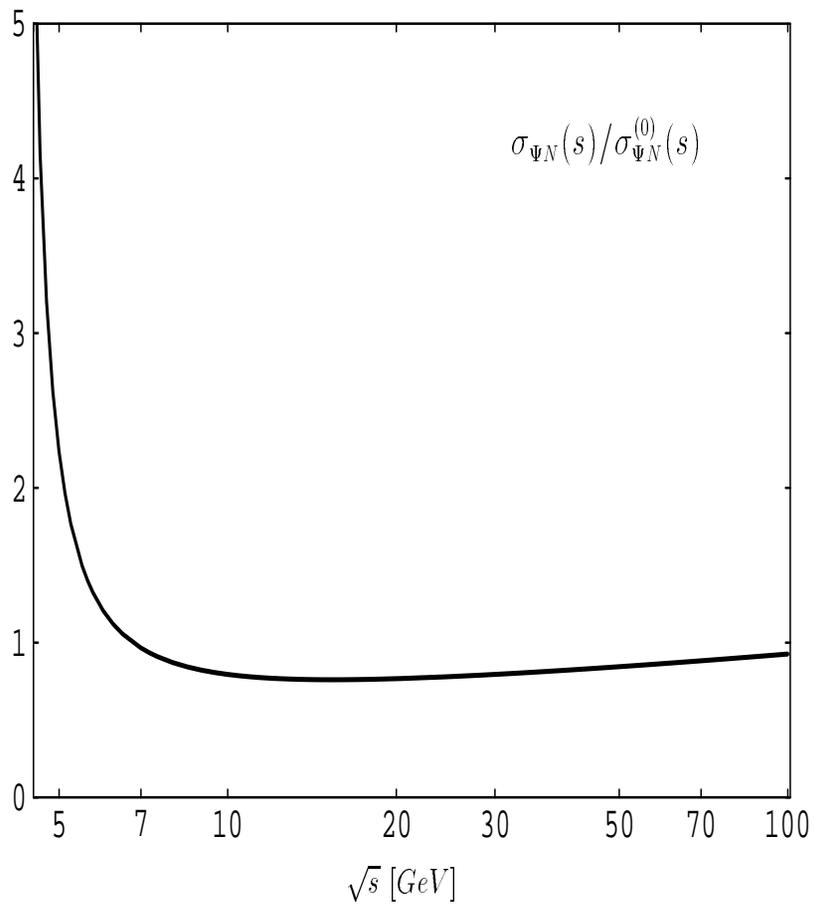
{\bf Fig.\ 1:} The ratio of
$\sigma_{J/\Psi-N}(s)$,
including target mass corrections, to the corresponding form
without these corrections; both forms are calculated with PDF (17).

\bigskip
\noindent
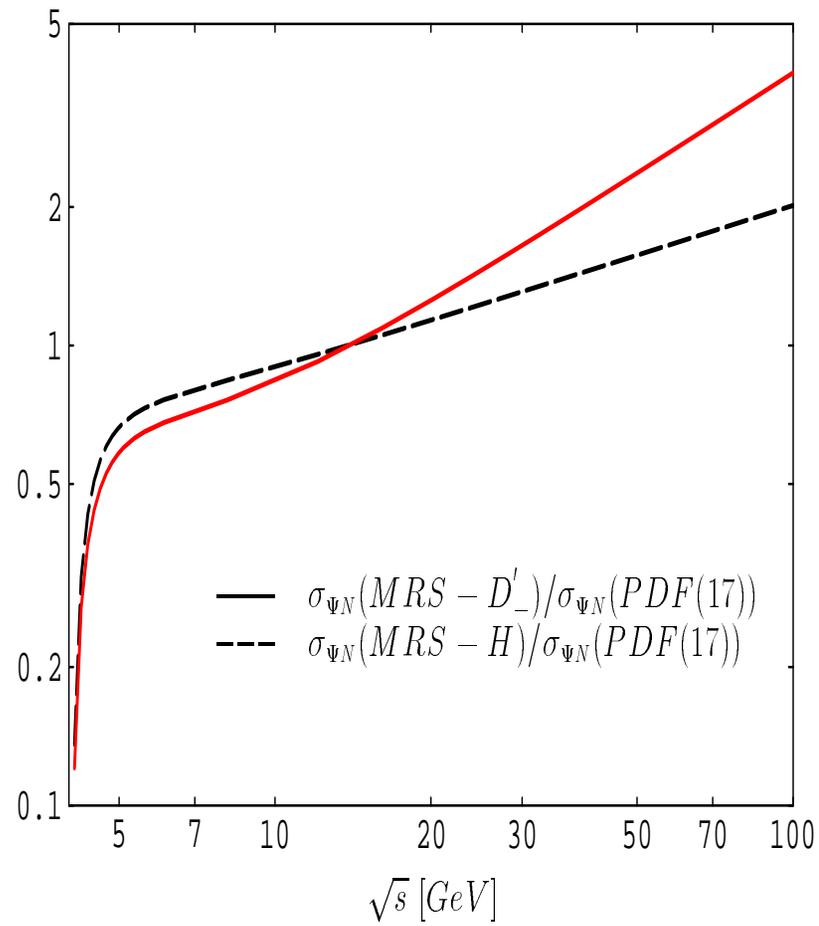
{\bf Fig.\ 2:} The ratio of $\sigma_{J/\Psi-N}(s)$, from
MRS D-' PDF (solid line) and from MRS H (dashed line), to the
cross section obtained from PDF (17); all cross sections include
target mass corrections.

\end{document}